\documentstyle[prd,aps,floats,psfig]  {revtex}

\def\lsim{\mathrel{\lower2.5pt\vbox{\lineskip=0pt\baselineskip=0pt
\hbox{$<$}\hbox{$\sim$}}}}
\def\gsim{\mathrel{\lower2.5pt\vbox{\lineskip=0pt\baselineskip=0pt
\hbox{$>$}\hbox{$\sim$}}}}

\newcommand{\ima}{{\mbox{Im}\,}}
\newcommand{\rea}{{\mbox{Re}\,}}
\newcommand{\be}{\begin{equation}}
\newcommand{\ee}{\end{equation}}

\newcommand{\NP}[1]{Nucl.\ Phys.\ {#1}}

\newcommand{\PL}[1]{Phys.\ Lett.\ {#1}}

\newcommand{\PR}[1]{Phys.\ Rev.\ {#1}}
\newcommand{\PRL}[1]{Phys.\ Rev.\ Lett.\ {#1}}

\begin{document}
\draft

\title{\LARGE \bf Unitarization of the complete meson-meson scattering
at one loop in Chiral Perturbation Theory.}


\author{\Large Jos\'e R. Pel\'aez and A. G\'omez Nicola}

\address{ \large 
  {Departamento de F\'{\i}sica Te\'orica II, Universidad Complutense,
28040 Madrid, SPAIN}}
\maketitle

\begin{abstract}
We report on our one-loop calculation of all the two meson
scattering amplitudes within SU(3) Chiral Perturbation Theory,
i.e. with  pions, kaons and etas. Once the amplitudes are
unitarized with the coupled channel Inverse Amplitude Method, they
satisfy simultaneously the correct low-energy chiral constraints
and unitarity. We obtain a remarkable description of meson-meson
scattering data up to 1.2 GeV including the scattering lengths and
 seven light resonances.
\end{abstract}

\maketitle

\subsection{Introduction}


 Chiral Perturbation Theory (ChPT) \cite{weinberg}
provides a powerful tool to describe the interactions of the
lightest mesons. These particles correspond to the Goldstone
bosons associated to the spontaneous breaking of the
$SU(3)_L\times SU(3)_R$ chiral symmetry down to $SU(3)_{L+R}$.
This would be  the symmetry breaking pattern of QCD if
the three lightest quarks were massless. Of course, quarks are not
massless, but since their masses are very small compared to the
typical hadronic scales, ${\cal O}$(1 GeV), their explicit
symmetry breaking effect only yields a small mass contribution for
the lightest mesons, which become pseudo-Goldstone bosons. Thus,
the three pions are the pseudo-Goldstone bosons of the $SU(2)$
spontaneous breaking when  only the $u$ and $d$ quarks are
considered. Similarly, when $s$ is also included, the eight
$SU(3)$  pseudo-Goldstone bosons  can be identified with the meson
octet formed by the pions,  kaons and the eta.

The low energy interactions of pions, kaons and the eta have to be
described with an effective Lagrangian respecting the above
described chiral  symmetry breaking pattern. Within ChPT,  only
pseudo-Goldstone bosons are included in the Lagrangian, thus
providing a low energy description. The possible terms compatible
with  the symmetry breaking pattern are organized in a derivative
and mass expansion (generically $p$). For instance, amplitudes are
obtained as an expansion in powers of the external momenta and the
quark masses over a typical chiral scale of ${\cal
O}$(1 GeV). One remarkable feature of the ChPT scheme is that all
loop divergences appearing at a given order in the expansion can
be absorbed by a finite number of (low energy) constants of the
counterterms that appear in the Lagrangian to the same order.
Therefore, order by order, the theory is finite and depends on a
few parameters, providing a predictive framework. Thus, once the
low-energy  constants are determined from just a few experiments,
predictions can be made for other processes.

This approach is very successful, but only at low energies
(usually, less than 500 MeV). For that reason, there is a growing
interest in developing methods to extend the ChPT applicability
range.  Among them,  the explicit introduction of heavier
resonances in the Lagrangian \cite{explicitresonances},
resummation of diagrams in a Lippmann-Schwinger or Bethe-Salpeter
approach \cite{LS}, or unitarization and dispersive techniques
like the Inverse Amplitude Method (IAM) \cite{Truong,IAM1}
applied to one-loop amplitudes. A version of the
latter, generalized to coupled channels provided a
remarkable description of meson-meson scattering up
to 1.2 GeV, generating dynamically seven
 light resonances \cite{IAM2}.

In principle, these methods respect the good low energy properties
of ChPT, since they are built from the perturbative results.
However, not all the meson-meson scattering processes had been
calculated at one loop in ChPT. The amplitudes available so far
are $\pi\pi\rightarrow\pi\pi$ \cite{Kpi}, $K\pi\rightarrow K\pi$
\cite{Kpi}, $\eta\pi\rightarrow\eta\pi$ \cite{Kpi} and the two
independent $K^+K^-\rightarrow K^+K^-$, $K^+K^-\rightarrow K^0
\bar{K}^0$ \cite{JAPaco}. Therefore, the IAM has only been applied
rigorously to the $\pi\pi$, $K \bar K$ final states, whereas for
the complete low-energy meson-meson scattering, additional
approximations had to be done \cite{IAM2}, meaning  in particular
that  it was not possible to compare with the low energy
parameters of standard ChPT in dimensional regularization or to
describe simultaneously the low and high energy regimes.

Here we report on our recent work \cite{nos} where we have
completed the calculation of the meson-meson scattering in
one-loop ChPT. There are three completely new amplitudes:
$K\eta\rightarrow K\eta$, $\eta\eta\rightarrow\eta\eta$ and
$K\pi\rightarrow K\eta$. In addition, we have recalculated
independently the other five amplitudes and all of them will be
given together in a unified notation, ensuring exact perturbative
unitarity and also correcting some misprints in the literature.

Once all the amplitudes are available,
we have done a coupled channel IAM  fit
 to describe the whole meson-meson scattering data below
1.2 GeV. Our results allow for a direct comparison with the
standard low-energy chiral parameters. Indeed, we find a very good
agreement with previous determinations from low-energy data using
standard ChPT. The main differences of our work with \cite{IAM2}
are that we consider the full one-loop calculation of the
amplitudes, which ensures their finiteness and scale independence
in dimensional regularization, we take into account the new
processes mentioned above and we are able to describe
simultaneously  the low energy and the resonance regions.

\subsection{The amplitudes}

The lowest order, ${\cal O}(p^2)$, meson-meson
scattering amplitudes (the low energy theorems) are obtained just
from the tree level diagrams of the lowest order Lagrangian. In
contrast, the calculation of the ${\cal O}(p^4)$ contribution
involves the evaluation of the following Feynman diagrams: First,
the one-loop diagrams in Fig.1, which are divergent.
 In particular those in Fig.1e, provide the wave function,
mass and decay constant  renormalizations, and that in
Fig.1a gives the imaginary part to ensure perturbative unitarity.  
Second, the tree level
graphs with the second order Lagrangian, which depend on the
chiral parameters $L_i$, that will absorb the previous
divergences through renormalization. In Table I, we list the
$L_i$ values from recent determinations. Note that the
parameters have been renormalized in the usual $\overline{MS}-1$
scheme of ChPT, using dimensional regularization. Thus, the
renormalized parameters have a scale dependence $L_i^r(\mu)$ (except
$L_3$ and $L_7$), and
they are given at $\mu=M_\rho$.

\begin{figure}
\centerline{\hbox{\psfig{file=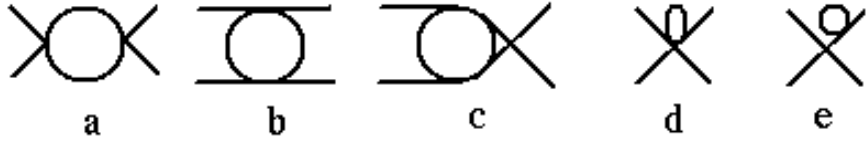,height=.08\textheight}}}

\vspace{.3cm}
\caption{Generic one-loop Feynman diagrams that have to be evaluated
in meson-meson scattering.}
\label{fig1:diagrams}
\end{figure}

\begin{table}[h]
\begin{tabular}{|c||c|c|c||c|}
\hline
 {Chiral Parameter} &
 {${\cal O}(p^6)$  $K_{l4}$ decays} &
 {${\cal O}(p^4)$  $K_{l4}$ decays} &
 {ChPT} &
 {IAM fits} \\
\hline
$L_1^r(M_\rho)$
& $0.53\pm0.25$
& $0.46$
& $0.4\pm0.3$& $0.56\pm0.10$ \\
$L_2^r(M_\rho)$
& $0.71\pm0.27$
& $1.49$
& $1.35\pm0.3$ & $1.21\pm0.10$\\
$L_3 $ & $-2.72\pm1.12$ & $-3.18$ & $-3.5\pm1.1$&
$-2.79\pm0.14$ \\
$L_4^r(M_\rho)$
& 0
& 0
& $-0.3\pm0.5$& $-0.36\pm0.17$ \\
$L_5^r(M_\rho)$
& $0.91\pm0.15$
& $1.46$
& $1.4\pm0.5$& $1.4\pm0.5$ \\
$L_6^r(M_\rho)$
& 0
& 0
& $-0.2\pm0.3$& $0.07\pm0.08$ \\
$L_7 $ & $-0.32\pm0.15$ & $-0.49$ & $-0.4\pm0.2$&
$-0.44\pm0.15$ \\
$L_8^r(M_\rho)$
& $0.62\pm0.2$
& $1.00$
& $0.9\pm0.3$& $0.78\pm0.18$ \\
\hline
\end{tabular}
\caption{Different sets of chiral parameters ($\times10^{3}$).
The first two columns come from recent analysis of $K_{l4}$ decays
at different orders [10] ($L_4$ and $L_6$ are set to
zero). In the third column $L_1,L_2,L_3$ come from
[11] and  the rest from  [1]. The last
one corresponds to the values from the IAM including the
uncertainty due to different systematic error used on different
fits.} 
\end{table}

After renormalization, the amplitudes  are finite and scale
independent. The details and results of the calculation
will be published elsewhere \cite{nos}. 
We will just recall that, in order to compare with experiment,
the amplitudes are projected into partial waves $t_{IJ}$ of
definite isospin $I$ and angular momentum $J$. Therefore, in the
chiral expansion we will have, omitting the $I,J$ subindices,
$t\simeq t_2+t_4+...$, where $t_2$ and $t_4$ the ${\cal
O}(p^2)$ and ${\cal O}(p^4)$ contributions, respectively.

\subsection{Unitarity}
The $S$ matrix unitarity relation  $S S^\dagger=1$
translates into simple relations for the elements of the $T$
matrix $t_{ij}$ if they are projected into partial waves,
where $i,j$ denote the different states physically available. For
instance, if there is only one possible state, "1",  the partial
wave $t_{11}$ satisfies
\begin{equation}
\ima t_{11}  = \sigma_1 \,\vert \,t_{11} \vert ^2 \quad
\Rightarrow \quad\ima \frac{1}{t_{11}}=-\sigma_1\quad\Rightarrow \quad t_{11} =
\frac{1}{\rea  t_{11} -i \,\sigma_1}
\label{uni1}
\end{equation}
where $\sigma_i=2 q_i/\sqrt{s}$ and $q_i$ is the C.M. momentum of
the state $i$. Written in this way it can be readily noted that
{\it we only need to know the real part of the Inverse Amplitude}.
The imaginary part is fixed by unitarity. In principle, this
relation {\it only holds above threshold} up to the energy where
another state, "2", is physically accessible. Above that point,
the unitarity relation for the partial waves can be written in
matrix form as:
\be
\ima T = T \, \Sigma \, T^* \quad \Rightarrow \quad \ima T^{-1}=- \Sigma
\quad  \Rightarrow \quad T=(\rea T- i \,\Sigma)^{-1}
\label{unimatrix}
\ee
with
\be
T=\left(
\begin{array}{cc}
t_{11}&t_{12}\\
t_{12}&t_{22} \\
\end{array}
\right)
\quad ,\quad
\Sigma=\left(
\begin{array}{cc}
\sigma_1&0\\
0 & \sigma_2\\
\end{array}
\right)\,,
\ee
which allows for an straightforward generalization to the
case of $n$ accessible states. Once more, unitarity
means that we would only need to
calculate the real part of the inverse amplitude matrix.
\begin{figure}
\hbox{\psfig{file=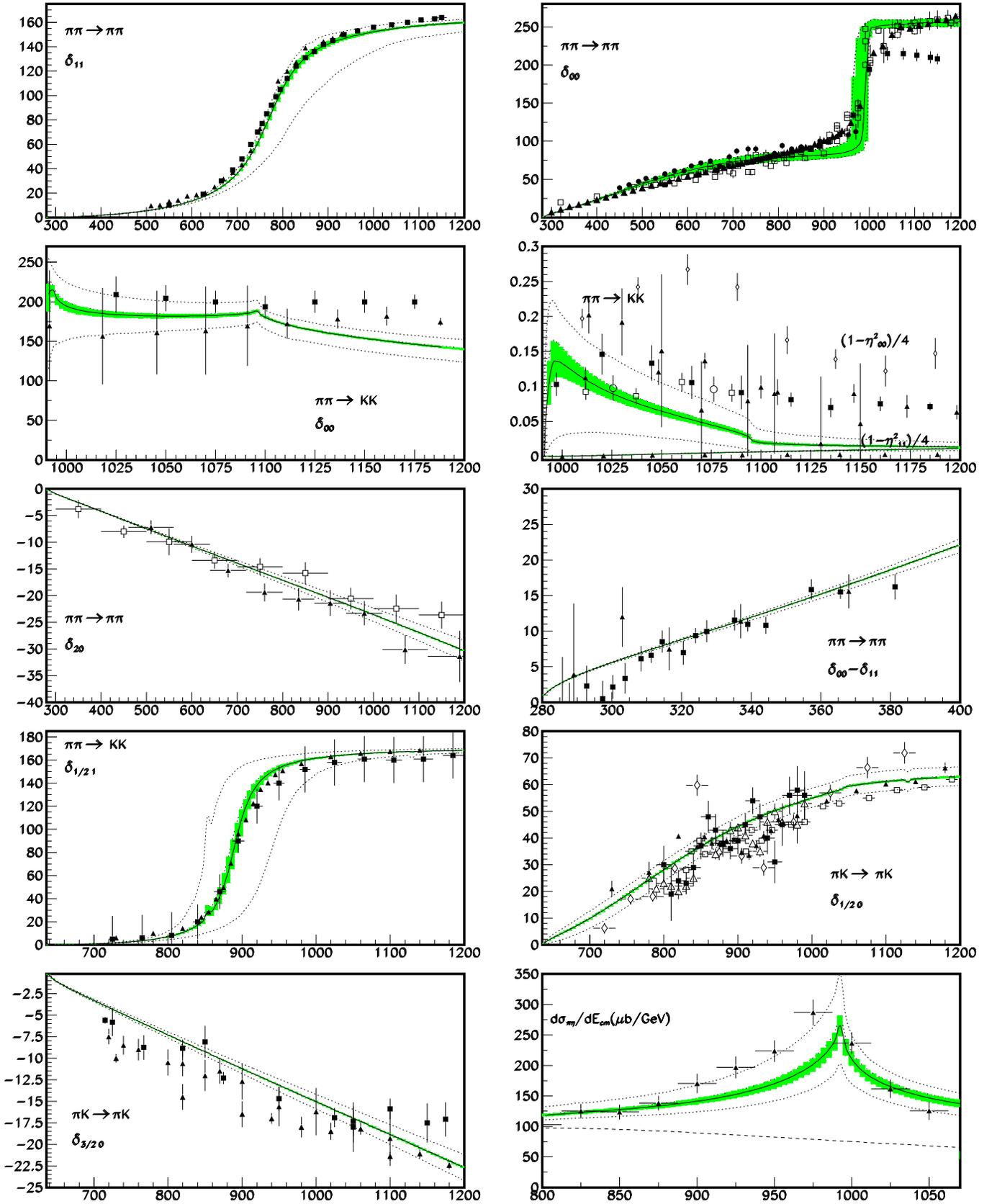,height=0.95\textheight}}

\vspace{.3cm}
  \caption{Result of the coupled channel
IAM fit to meson-meson scattering data (see [9] for references).
The shaded area covers the uncertainty due to MINUIT errors.
The area between the dotted lines corresponds to the
uncertainty in the $L^r_i$  due to the
use of different systematic errors on the fits. The dashed line in the last plot is the continuous background underneath the resonant contribution.}
\end{figure}

Note that the above unitarity relations are non-linear. This
implies that they will never be satisfied exactly with polynomials
like the amplitudes obtained from ChPT. Nevertheless,
unitarity holds perturbatively, i.e,
\begin{eqnarray}
\ima T_2 = 0+ {\cal O}(p^4), \quad \quad \ima T_4 = T_2 \, \Sigma
\, T_2^*\,+ {\cal O}(p^6) , \label{pertuni}
\end{eqnarray}
\subsection{Unitarization: The inverse Amplitude Method}
One of the simplest methods to unitarize the chiral amplitudes is to
introduce the $\rea T$ in eq.(\ref{unimatrix}), calculated as a ChPT expansion
\begin{eqnarray}
  T^{-1}&\simeq& T_2^{-1}(1-T_4 T_2^{-1}+...),\\
\rea  T^{-1} &\simeq&  T_2^{-1}(1-(\rea T_4) T_2^{-1}+...),
\end{eqnarray}
Taking into account the perturbative unitarity conditions,
 eq.(\ref{pertuni}),
we find
\begin{equation}
 T\simeq T_2 (T_2-T_4)^{-1} T_2,
\label{IAM}
\end{equation}
which is the coupled channel Inverse Amplitude Method, which will
use to unitarize simultaneously all the one-loop ChPT meson-meson
scattering amplitudes. This method is able to generate seven
resonant states. The novelty of our approach is that, since we
have the complete ${\cal O}(p^4)$ ChPT amplitudes, we can
simultaneously recover the very same ChPT amplitudes up to ${\cal
O} (p^4)$, and thus have a good low energy limit. 
\subsection{Results and conclusion}
We can now use previous determinations of the
chiral parameters with the IAM  and even the correct resonant behavior
resonances. Once more, we can use the $L_i^r$ because we have the
complete amplitudes renormalized in the $\bar{MS}-1$ scheme. We
have nevertheless carried out a fit (using MINUIT \cite{MINUIT})
of the presently available data on meson-meson scattering.  Since
there are incompatibilities between different experiments,
customarily a $1\%$, $3\%$ and $5\%$ systematic error has been
added, which introduces an additional source of error. We give in
Table 1 the resulting chiral parameters from the fit, whose errors
correspond to those of MINUIT combined with those from the
systematic uncertainty. Note that they are compatible 
with previous determinations.

In Fig.2 we show the results of the IAM fit to these data, which
is given in terms of phase shifts, inelasticities, and mass
distributions of different processes (see \cite{nos} for details).
The gray error bands cover the uncertainties in the $L_i$ 
due to MINUIT, and are calculated by a Monte-Carlo 
gaussian sampling of the parameters. The area between the
dotted lines has been calculated similarly but with the errors
in the chiral parameters due to the different choice of systematic error.
It can be noticed that all the resonant features are
reproduced. However, thanks to the new amplitudes we are also able
to obtain simultaneously values for the threshold parameters (they
have not been fitted) which are listed in table 2. Note the good
agreement with the experimental values when they exist.

\begin{figure}[h]
\hbox{\psfig{file=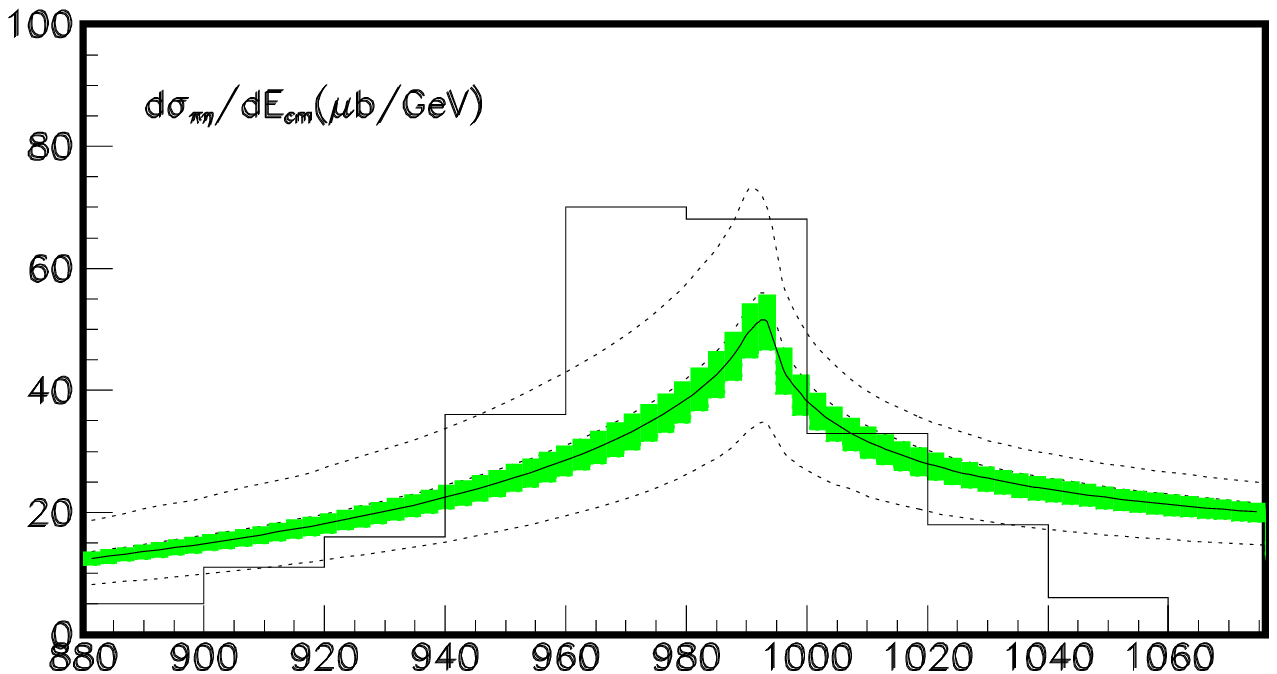,width=9cm}
\psfig{file=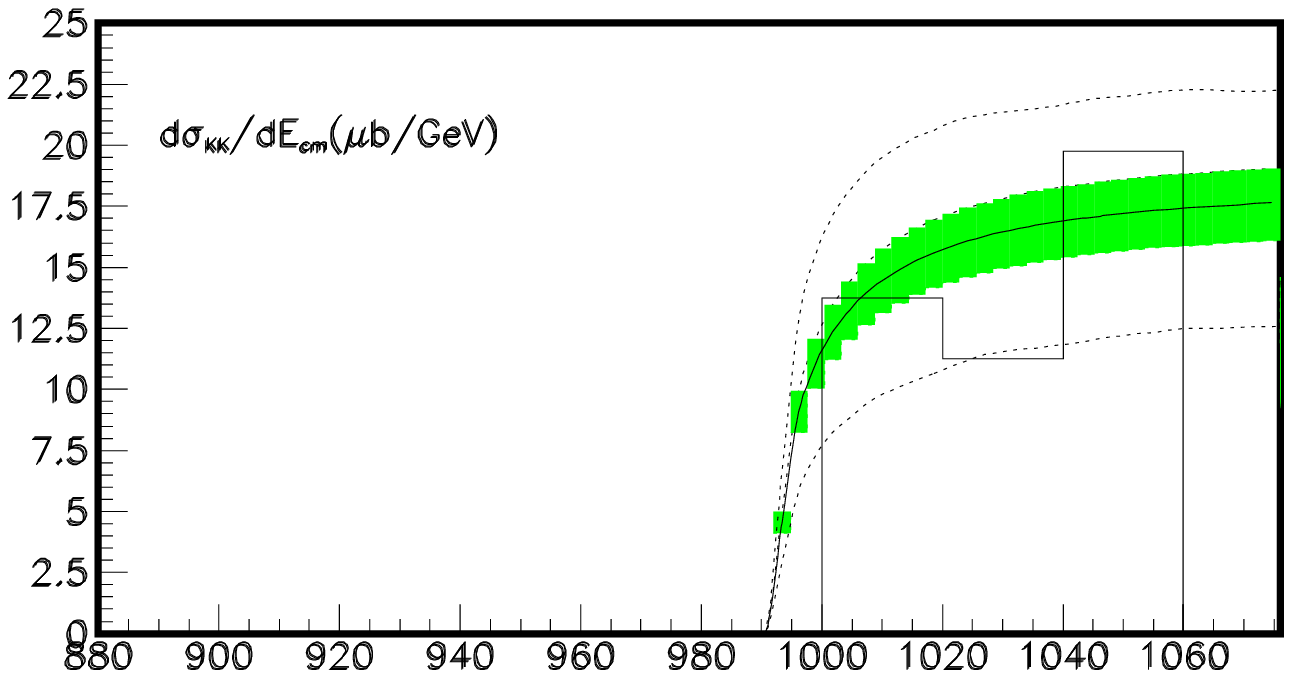,width=9cm}}
\caption{ Effective mass distributions of the two
mesons in the final state of
$K^-p\rightarrow\Sigma^+(1385)\pi\eta$ and
$K^-p\rightarrow\Sigma^+(1385)K \bar{K}$. This plots are not part of the IAM fit. For
data references see [9].}
\label{fig3:flatte}
\end{figure}

\begin{table}[h]
\begin{tabular}{|c|c|c|c|c|}
\hline 
Threshold&Experiment&IAM fit&ChPT ${\cal O}(p^4)$&ChPT ${\cal O}(p^6)$\\
parameter&&&\cite{explicitresonances,IAM1,Kpi}&\cite{bij}\\ 
\hline \hline 
$a_{0\,0}$&0.26 $\pm$0.05&0.231$^{+0.003}_{-0.006}$&0.20&0.219$\pm$0.005\\
$b_{0\,0}$&0.25 $\pm$0.03&0.30$\pm$ 0.01&0.26&0.279$\pm$0.011\\
$a_{2\,0}$&-0.028$\pm$0.012&-0.0411$^{+0.0009}_{-0.001}$&-0.042&-0.042$\pm$0.01\\
$b_{2\,0}$&-0.082$\pm$0.008&-0.074$\pm$0.001&-0.070&-0.0756$\pm$0.0021\\
$a_{1\,1}$&0.038$\pm$0.002&0.0377$\pm$0.0007&0.037&0.0378$\pm$0.0021\\ 
$a_{1/2\,0}$&0.13...0.24&0.11$^{+0.06}_{-0.09}$&0.17&\\
$a_{3/2\,0}$&-0.13...-0.05&-0.049$^{+0.002}_{-0.003}$&-0.5&\\
$a_{1/2\,1}$&0.017...0.018&0.016$\pm$0.002&0.014&\\
$a_{1\,0}$&&0.15$^{+0.07}_{-0.11}$&0.0072&\\ 
\hline
\end{tabular}
\vspace{.3cm}
\caption{ Scattering lengths $a_{I\,J}$ and slope parameters
$b_{I\,J}$ for different meson-meson scattering channels. For
experimental references see [9]. Let us
remark that our one-loop IAM results are very similar
to those of two-loop ChPT.}
\end{table}

\section*{Acknowledgments} Work  partially supported from the Spanish
CICYT projects AEN97-1693, FPA2000-0956, PB98-0782 and
BFM2000-1326.





\end{document}